\documentclass{emulateapj} 
\usepackage{apjfonts} 
\slugcomment{{\sc \apj\ ACCEPTED [26 July 2012]}} 
\newcommand{\CO}{$^{12}$CO} 
\newcommand{\bandhead}{CO$_{2.29}$}
\newcommand{\wave}{2.29~$\mu$m} 
\newcommand{\ew}{W$_{2.29}$}
\newcommand{\tna}{\tablenotemark{$\dagger$}}

\begin{document}

\title{\sc Measurement of the Mass and Stellar Population Distribution in M82 with the LBT}

\author{Johnny P. Greco, Paul Martini, and Todd A. Thompson} 
\affil{Department of Astronomy,
The Ohio State University, Columbus, OH 43210, USA;
greco.40@buckeyemail.osu.edu} 
\shorttitle{THE MASS \& STELLAR
POPULATION DISTRIBUTION IN M82} 
\shortauthors{GRECO, MARTINI, \& THOMPSON}

\begin{abstract}

We present a $K-$band spectroscopic study of the stellar and gas kinematics,
mass distribution, and stellar populations of the archetypical starburst galaxy
M82. Our results are based on a single spectrum at a position angle of
$67.5\arcdeg\!$ through the $K-$band nucleus. We used the \CO\ stellar
absorption band head at \wave\ (\bandhead) to measure the rotation curve out to
nearly 4~kpc radius on both the eastern and western sides of the galaxy.  Our
data show that the rotation curve is flat from 1~$-$~4~kpc. This stands in
sharp contrast to some previous studies, which have interpreted H~$\!$I and CO
emission-line position-velocity diagrams as evidence for a declining rotation
curve. The kinematics of the Br$\gamma$, H$_2$, and He~I emission lines are
consistent with, although characterized by slightly higher velocities than, the
stellar kinematics. We derived M82's mass distribution from our stellar
kinematic measurements and estimate its total dynamical mass is
$\sim$10$^{10}~M_{\odot}$. We measured the equivalent width of \bandhead\ (\ew)
as a function of distance from the center of the galaxy to investigate the
spatial extent of the red supergiant (RSG) population. The variation in \ew\
with radius clearly shows that RSGs dominate the light inside 500~pc radius.
M82's superwind is likely launched from this region, where we estimate the
enclosed mass is $\lesssim$~2$\times 10^{9}~M_{\odot}$. 

\end{abstract}


\section{Introduction} \label{sec:introduction}

Observations suggest that star formation in galaxies is regulated by feedback 
processes. In rapidly star-forming galaxies, feedback operates on large scales, 
driving galaxy-wide superwinds, which can eject the raw materials for future 
stellar generations if the interstellar material is launched with enough 
momentum to escape the host galaxy's gravitational potential. Otherwise, 
the ejected material may be re-accreted in a so-called ``fountain flow''. These 
issues are crucial for understanding the formation of stars in rapidly star-forming
galaxies across cosmic time.

The exceptionally high star formation rates present in starburst galaxies make
them optimal targets to probe feedback processes. The focus of this paper is
the archetype of this class of objects, M82. This galaxy's  proximity
\citep[3.63 Mpc; $1''=17.6$pc;][]{Fr94, Sa99, Ge11} and nearly edge-on geometry
\citep[$i=80\arcdeg$;][]{LS63,Mc95} make it a superb laboratory for studying the
physics of galactic winds and star formation.  At infrared wavelengths, its
luminosity is $6 \times 10^{10}~L_{\odot}$ \citep{San03}, which makes it one of
the brightest and most observed infrared objects in the sky. Most of its
infrared luminosity originates in the starburst core, which is severely
obscured by dust at visible and ultraviolet wavelengths \citep{Ri80, Fo01}.

The origin of M82's starburst is thought to be a tidal interaction with its
more massive companion, M81, $\sim$10$^8$ years ago \citep{Got77, Lo87, Te91,
Ac95}.  The pronounced H~$\!$I bridge that connects the galaxies
\citep{Co77,Yu93} is strong evidence for such an interaction, and \citet{Yu94}
found that the ``cloud of H~$\!$I'' that engulfs the M81 group is dominated by
filamentary tidal structures, which suggests a violent past of tidal
interaction. In addition to the dynamical effects associated with tidal forces
from M81, the global dynamics and evolution of the starburst are further
perturbed by the presence of a stellar bar \citep{Te91}, which may have formed
as a result of the interaction. 

As an interacting, barred galaxy, M82's observed dynamical properties are
distinct from those seen in typical Sb and Sc galaxies.  Molecular gas
measurements by \citet{Yo84} show that the CO velocity field contains warps and
peculiarities that are consistent with a significant non-circular velocity
component, perhaps due to radial infall or a polar ring tilted with respect to
the major axis. In addition, measurements of CO (2,1) by \citet{Sofue92}
suggest that the rotation of the gas disk is nearly Keplerian at radii $>$
1~kpc. \citet{So98} reproduced this Keplerian motion with a model in which the
outer disk and halo were stripped during the encounter with M81. Although flat
rotation curves are observed in the vast majority of galaxies, \citet{RuFord83}
found falling rotation curves in several tidally interacting galaxies, which
suggests that some dark matter is stripped from galaxies during interactions. 

The H~$\!$I velocity field along the major axis is similar to the CO velocity
field; it also has peculiarities such as a sudden drop at $\sim$1~kpc, which
some authors have interpreted as a falling rotation curve \citep{Yu93}.  It is
important to note, however, that the H~$\!$I gas is highly disrupted and may
not provide a reliable rotation curve for the outer disk. In fact, \citet{Yu93}
state that rotation dominates the H~$\!$I kinematic signature inside the
central 1~kpc, but the outer region is characterized by a significant, likely
tidally induced, nonrotational component.  

The gas and stars in M82 have been shown to be composed of two distinct
kinematic structures. \citet{West09} found that the axis of gas rotation is
offset from the stellar rotation axis and the photometric major axis by
$\sim$12$^\circ$.  This result implies that, in long-slit spectroscopic
studies, the shape of the rotation curve is particularly sensitive to the
position and alignment of the slit. In addition, \citet{Kont09} studied M82's
star cluster population out to $\sim$2.6~kpc radius and found that the rotation
curve as traced by the clusters is flat at large radii, which further
distinguishes the stellar kinematic tracers from the previous measurements of
the gas kinematics. 

\citet{Fo03} proposed that M82's starburst activity occurred in two successive
episodes that each lasted a few million years and peaked about 10$^7$ and
5$\times$10$^6$ years ago. In this model, the encounter with M81 triggered the
first burst in the central 500~pc by inducing strong large-scale torques, loss
of angular momentum, and an infall of molecular clouds toward the nuclear
region. The second burst occurred in a circumnuclear ring and along the stellar
bar and is attributed to bar-induced dynamical resonances. The \citet{Fo03}
starburst model can explain the apparent abundance of red supergiants (RSGs)
that dominate the light within 500~pc radius.

In order to understand the dynamics of the ejected gas and constrain the
physics of feedback processes in M82, it is essential to accurately measure the
total dynamical mass on both the small scales of star-forming regions and the
large scales of the entire galaxy. Rotation curves are most commonly employed
to determine the distribution of mass in galaxies \citep{So01}, and many
studies have employed gas and stellar kinematics at a range of wavelengths to
measure radial velocities in M82 \citep[e.g.,][]{Saito84, Go90, Mc93, Ac95,
Sofue92}.  \citet{Gr11} includes a compilation of these results. Given that M82
has recently undergone tidal interactions and harbors an extraordinary
superwind, stellar kinematics should be a more reliable tracer of the mass
distribution than the gas. Nearly all previous stellar kinematic measurements
used visible-wavelength tracers, which may be compromised by the substantial
dust attenuation present in the galaxy.  The near-infrared $K-$band is a better
tracer of the stellar kinematics, as attenuation $A_{\lambda}$ in the $K-$band
is approximately a factor of 10 less than in the $V-$band. In this study, we
use $K-$band spectroscopy to study M82's gas and stellar kinematics, mass
distribution, and stellar populations.

\section{Observations and Data Reduction} \label{sec:reduce}

\begin{figure}[h] 
\centering 
\includegraphics[width=7.2cm,height=5.8cm, trim=0cm 0cm 0cm 1cm,clip=true]{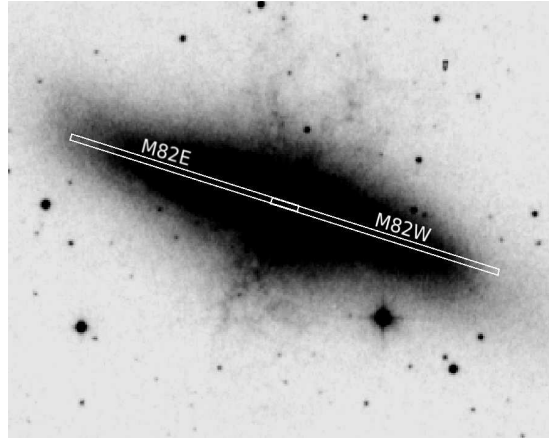} 
\caption{ 
Digital Sky Survey image of M82 with overlays of our two slit positions. Each
slit is $4'$ long and they overlap on the center of the galaxy. The widths of
the slits have been increased by a factor of six for ease of visibility in this
figure. North is up and East is to the left.
}
\label{fig:slits} 
\end{figure}

We obtained $K-$band spectra of M82 at the Mount Graham International
Observatory with the LUCI-1 spectrograph and the Large Binocular Telescope
(LBT) on 2011 February 10. The observations were made through a $K-$band filter
(2.05\,-\,2.37\,$\mu$m) with the N1.8 camera, which has a plate scale of
$0''\!\!.25$ pixel$^{-1}$ and a field of view of $4'\!\times4'$. A grating with
210 lines mm$^{-1}$ was used with a slit width of $1''\!$. This combination
yields a spectral resolution of $\lambda / \Delta\lambda\approx$\,3000 or 100
km~s$^{-1}$. With sufficient signal-to-noise, the centroid of the
cross correlation function for relative velocity measurements can be measured
with much higher resolution. Two slit positions, which we designate as M82E and
M82W, were used to measure the rotation curve out to $4'\,(\sim$4\,kpc) from
the center of the galaxy. We centered each end of the slit on the nuclear
region at a position angle of $67.5\arcdeg\!$ (Figure~\ref{fig:slits}). The
observations were made with 300 second exposures in an {\it object-sky}
sequence for a total on-source integration time of 1 hour per slit position.  A
telluric standard close in air mass to M82 was observed before and after both
the M82E and M82W observations.

We carried out standard data reduction with the IRAF\footnote{IRAF is
distributed by the National Optical Astronomy Observatory, which is operated by
the Association of Universities for Research in Astronomy, Inc., under
cooperative agreement with the National Science Foundation.}\,software package.
The two-dimensional data were transformed onto an orthogonal
wavelength/slit-position grid based on a wavelength solution from the sky's
bright OH emission lines. This transformation linearized the dispersion to
$\sim$1.64~\AA\ pixel$^{-1}$. We used the {\tt xtellcor} \citep{Va03} package
in IDL to correct for absorption from Earth's atmosphere. The telluric
standard spectra had slightly higher spectral resolution than was obtained for
M82. This is likely because the light from the stars did not fill the 1$''$
wide slit. This complication was remedied by convolving the telluric spectra
with Gaussians to match their resolutions to M82.

\begin{figure}[h]
\center
\includegraphics[width=9.4cm,trim=0cm 0.8cm 0cm 2cm,clip=true]{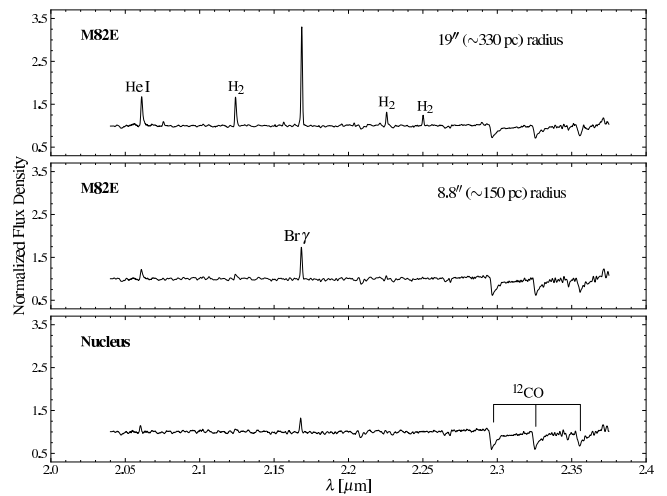}
\caption{
$K-$band spectra from M82's nucleus, 8.8$''$ east from center, and
19$''$ east from center. \bandhead, which is the dominant absorption
feature, originates in the atmospheres of cool, evolved stars. The He$\!$~I and
Br$\gamma$ lines trace star formation, and the H$_2$ is evidence of shocked
gas. The strength of the emission lines peaks near 19$''$. 
}
\label{fig:spec}
\end{figure}
\begin{figure*}[!t] 
\centering                     
\includegraphics[width=19.5cm,trim=2cm 1cm 0cm 1cm,clip=true]{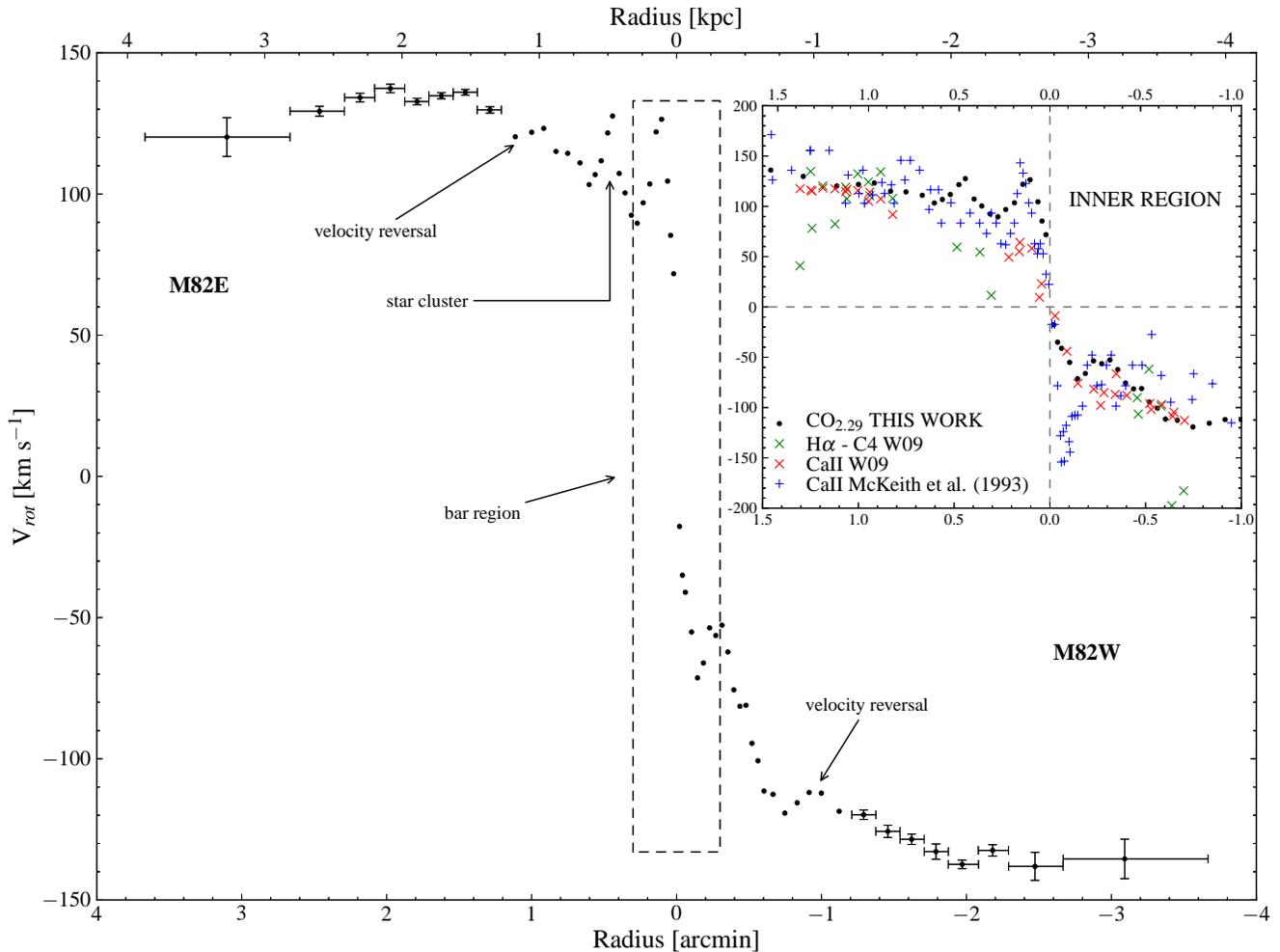}
\caption{
Stellar rotation curve based on \bandhead.  The horizontal error bars indicate
the aperture size of the measurement, and the vertical error bars represent the
rms scatter between the six cross correlation templates. All data points
without error bars have uncertainties less than 2 km~s$^{-1}$ and aperture
sizes less than 5$''$.  We assume an inclination of 80$^\circ$ for all velocity
measurements. The symmetric, sharp features on either side of the nucleus are
due to the non-circular motions of the stellar bar. At $\sim$1~kpc on either
side of the galaxy, there is an apparent ``velocity reversal'', which may be
due to symmetric spiral arms. The inset compares our results of the inner
region with stellar velocity measurements by \citet[W09;][]{West09} and
\citet{Mc93}. The C4 label with the W09 data refers to the component of an 
integral field unit (IFU) observation. 
} 
\label{fig:CO} 
\end{figure*}

\vspace{-.4cm}
\section{Rotation Curve} 

To measure the rotation curve, we symmetrically extracted 30 apertures each
from the M82E and M82W slit positions. We progressively increased the aperture
widths at larger radii to achieve a sufficient signal-to-noise ratio in the
stellar continuum. Figure~\ref{fig:spec} shows illustrative $K-$band spectra
from M82's nucleus and eastern side. We measured the rotation curve out to
nearly 4~kpc radius with the \CO\ stellar absorption band head at \wave\
(\bandhead) and to $\sim$1~kpc radius with gas emission (He~I, H$_2$,
and Br$\gamma$) from the interstellar medium. We have assumed an inclination of
80$^\circ$ and corrected each velocity accordingly. It is important to note
that the choice of dynamical center may introduce systematic uncertainties, as
the photometric center does not necessarily coincide with the dynamical center.
As the gas and stars trace distinct kinematic structures, we discuss each in
turn. 


\begin{deluxetable*}{rcrr || rcrr}[!h]
\centering
\tablecolumns{8}
\tablewidth{0pt}
\tabletypesize{\scriptsize}
\tablewidth{6.0truein}
\tablecaption{Stellar Rotation Curve Data\label{tbl:curve}}
\tablehead{
\multicolumn{4}{c}{\bf M82E}  &  \multicolumn{4}{c}{\bf M82W} \\ \hline \\ 
\colhead{Position} &
\colhead{Aperture} &
\colhead{Velocity} &
\colhead{Uncertainty} &
\colhead{Position} &
\colhead{Aperture} &
\colhead{Velocity} &
\colhead{Uncertainty} \\
\colhead{(arcmin)} &
\colhead{(arcmin)} &
\colhead{(km s$^{-1}$)} &
\colhead{(km s$^{-1}$)} &
\colhead{(arcmin)} &
\colhead{-1$\times$(arcmin)} &
\colhead{(km s$^{-1}$)} &
\colhead{(km s$^{-1}$)} 
}
\startdata
0.0203  &      0.0125 - 0.0292     & 71.8   & 1.8  & -0.0205  &      0.0125 - 0.0292     &  -17.7   &  1.3  \\
0.0411  &\,\,\,0.0333 - 0.0500\tna & 85.4   & 1.9  & -0.0403  &\,\,\,0.0333 - 0.0500\tna &  -35.0   &  1.2  \\
0.0622  &      0.0417 - 0.0833     & 104.6  & 1.7  & -0.0609  &      0.0417 - 0.0833     &  -41.0   &  1.2  \\
0.1034  &      0.0833 - 0.1250     & 126.5  & 1.2  & -0.1031  &      0.0833 - 0.1250     &  -55.1   &  1.0  \\
0.1412  &      0.1250 - 0.1667     & 122.0  & 1.2  & -0.1443  &      0.1250 - 0.1667     &  -71.4   &  0.8  \\
0.1848  &      0.1667 - 0.2083     & 103.6  & 1.7  & -0.1851  &      0.1667 - 0.2083     &  -66.1   &  2.0  \\
0.2303  &      0.2083 - 0.2500     & 96.9   & 1.4  & -0.2278  &      0.2083 - 0.2500     &  -53.6   &  2.7  \\
0.2711  &      0.2500 - 0.2917     & 89.7   & 1.3  & -0.2706  &      0.2500 - 0.2917     &  -56.4   &  2.2  \\
0.3121  &      0.2917 - 0.3333     & 92.5   & 1.0  & -0.3133  &      0.2917 - 0.3333     &  -52.7   &  1.7  \\
0.3553  &      0.3333 - 0.3750     & 100.4  & 1.0  & -0.3542  &      0.3333 - 0.3750     &  -62.2   &  1.8  \\
0.3955  &      0.3750 - 0.4167     & 107.3  & 1.3  & -0.3950  &      0.3750 - 0.4167     &  -75.6   &  1.7  \\
0.4413  &      0.4166 - 0.4583     & 127.6  & 1.2  & -0.4372  &      0.4166 - 0.4583     &  -81.4   &  1.6  \\
0.4747  &      0.4583 - 0.5000     & 121.6  & 1.0  & -0.4789  &      0.4583 - 0.5000     &  -81.0   &  1.7  \\
0.5198  &      0.5000 - 0.5417     & 111.8  & 1.2  & -0.5200  &      0.5000 - 0.5417     &  -94.5   &  1.4  \\
0.5614  &      0.5416 - 0.5833     & 106.9  & 1.3  & -0.5615  &      0.5416 - 0.5833     &  -100.7  &  1.4  \\
0.6032  &\,\,\,0.5833 - 0.6250\tna & 103.3  & 1.8  & -0.6037  &\,\,\,0.5833 - 0.6250\tna &  -111.4  &  1.9  \\
0.6662  &      0.6250 - 0.7083     & 111.0  & 1.5  & -0.6657  &      0.6250 - 0.7083     &  -112.6  &  1.7  \\
0.7512  &      0.7083 - 0.7917     & 114.4  & 1.5  & -0.7471  &      0.7083 - 0.7917     &  -119.3  &  1.4  \\
0.8318  &      0.7916 - 0.8750     & 115.1  & 1.8  & -0.8325  &      0.7916 - 0.8750     &  -115.6  &  1.7  \\
0.9166  &\,\,\,0.8750 - 0.9583\tna & 123.2  & 1.6  & -0.9149  &\,\,\,0.8750 - 0.9583\tna &  -111.9  &  1.9  \\
0.9994  &      0.9583 - 1.0416     & 121.9  & 1.3  & -0.9990  &      0.9583 - 1.0416     &  -112.2  &  1.8  \\
1.1127  &      1.0416 - 1.2083     & 120.3  & 1.2  & -1.1217  &      1.0416 - 1.2083     &  -118.6  &  1.8  \\
1.2876  &      1.2083 - 1.3750     & 129.8  & 1.1  & -1.2908  &      1.2083 - 1.3750     &  -119.8  &  1.7  \\
1.4572  &      1.3750 - 1.5417     & 136.0  & 1.0  & -1.4576  &      1.3750 - 1.5417     &  -125.7  &  2.2  \\
1.6221  &      1.5417 - 1.7083     & 134.8  & 1.1  & -1.6220  &      1.5417 - 1.7083     &  -128.5  &  1.8  \\
1.7900  &      1.7083 - 1.8750     & 132.7  & 1.1  & -1.7897  &      1.7083 - 1.8750     &  -132.9  &  2.7  \\
1.9741  &      1.8750 - 2.0833     & 137.4  & 1.5  & -1.9705  &      1.8750 - 2.0833     &  -137.4  &  1.5  \\
2.1838  &      2.0833 - 2.2917     & 134.1  & 1.5  & -2.1801  &      2.0833 - 2.2917     &  -132.4  &  2.0  \\
2.4630  &      2.2917 - 2.6667     & 129.3  & 1.8  & -2.4728  &      2.2917 - 2.6667     &  -138.1  &  5.0  \\
3.1024  &      2.6667 - 3.6667     & 120.2  & 6.9  & -3.0919  &      2.6667 - 3.6667     &  -135.5  &  7.0  
\enddata
\tablecomments{
All values are measured with respect to M82's center. The positions are flux weighted, and the uncertainties
were calculated as the rms scatter between the six templates. At M82's distance, $1'\approx 1.056$~kpc.
\tablenotetext{$\dagger$}{Cross-correlation aperture}
}
\end{deluxetable*}

\subsection{Stellar Absorption}

\bandhead\ originates in the atmospheres of cool, evolved stars and provides a
strong absorption feature for stellar velocity measurements
\citep[e.g.,][]{Ga95}. Previous studies of M82 have used H$\alpha$ and NaD to
measure the rotation curve with stellar kinematics \citep[e.g.][]{Ma60,Go90},
yet these measurements are likely compromised by the $A_V\,\sim5 \rightarrow25$
attenuation toward the central region \citep{Ri80, Le90, Pu91}. At somewhat
longer wavelengths, \citet{Mc93} measured the stellar kinematics  with Ca II
stellar absorption lines (8662, 8542, and 8498~\AA).  If the near-infrared
extinction wavelength dependence is given by the relation $A_{\lambda} \propto
\lambda^{-1.75}$ \citep{Dr89}, attenuation for \bandhead\ is approximately a
factor of five less than Ca II. With the copious amount of gas and dust
present, an improvement at this level may reveal  kinematic features
that are otherwise obscured. 

To measure the stellar rotation curve, we selected six apertures with high
signal-to-noise as cross-correlation templates for relative velocity
measurements. We used the {\tt rvsao} package in IRAF to cross-correlate all
60 apertures against each of the six templates. We restricted the wavelength
range from 2.28 to 2.365$\mu$m to focus the analysis on \bandhead.  This
resulted in six independent rotation curves. We assumed the galaxy is symmetric
on large scales to determine zero velocity. After correcting for M82's nearly
edge-on inclination, we took the mean velocity at each point to be the value of
the rotation curve at that point, and the uncertainty was calculated as the rms
scatter between the six curves.  Figure~\ref{fig:CO} shows our stellar rotation
curve.  The horizontal bars represent aperture size.  All data points without
error bars have uncertainties less than 2~km~s$^{-1}$ and aperture sizes less
than 5$^{\prime \prime}$. Table \ref{tbl:curve} lists the measurements
(apertures selected as cross-correlation templates are marked with a dagger).

The observed stellar velocities show the stellar rotation curve is flat on
scales of 1~$-$~4~kpc. This seems to contradict the falling rotation curve
suggested by H~I and CO emission from the interstellar medium
\citep{Sofue92,Yu93}. The gas, however, is severely disrupted by the wind and
may also be affected by radial inflow and tides \citep{Yo84}. The stellar data
are therefore expected to trace the mass distribution much more reliably. 

The rotation curve for M82E appears to decline slightly from $\sim 2
\rightarrow 4$~kpc. This decline has $\sim$2$\sigma$ significance and may be
due to tidal stripping caused by the interaction with M81. The decline in the
rotation curve is at the approximate location of a flaring in the stellar disk
and the northern H~I streamer \citep{Yu93}. No decline is seen in the M82W
data. Measurements at larger radii are needed to determine whether or not the
falloff is due to tidal stripping.

\begin{figure*}[!t] 
\center 
\includegraphics[width=20.0cm,trim=5.4cm .4cm 0cm 0cm,clip=true]{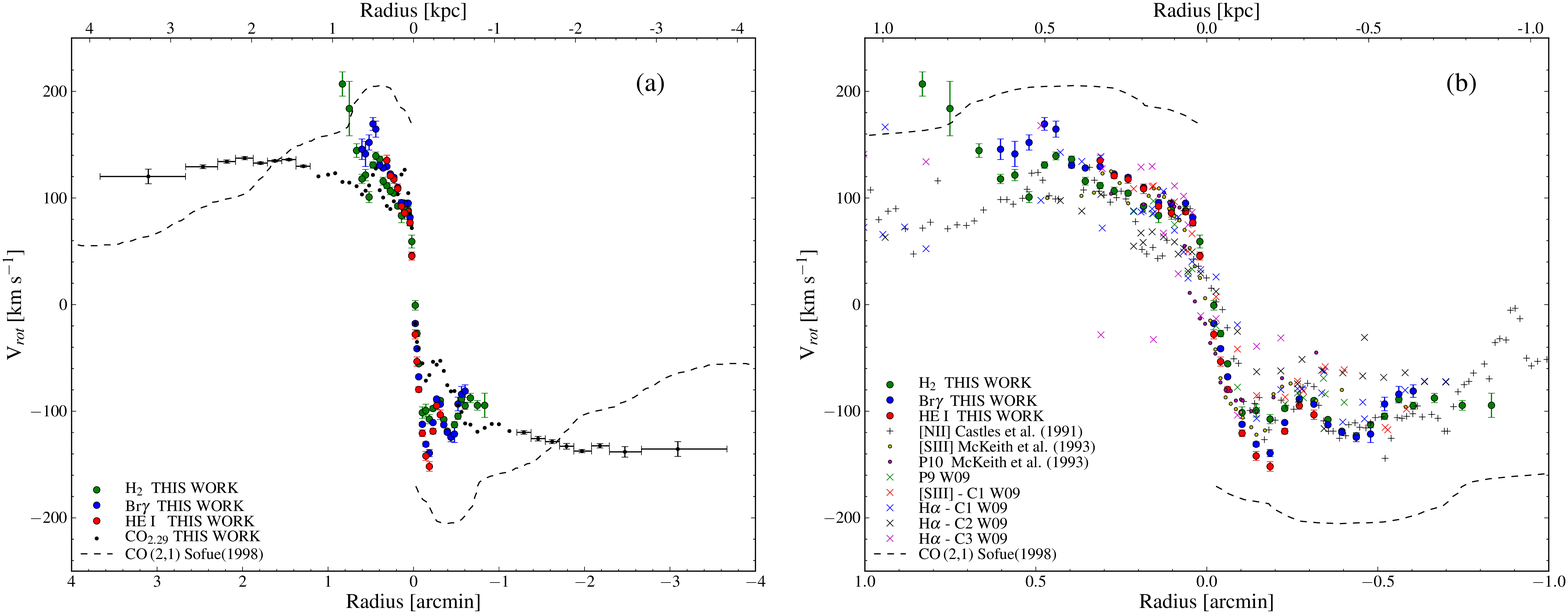}
\caption{
(a) The gas and stellar rotation curves from this work and the measurements of
CO (2,1) by \citet{So98}. On average, the gas has larger velocities than the
stars. (b) Comparison of gas measurements of the inner 1~kpc radius from this
work with gas measurements by \citet{Ca91}, \citet{Mc93},
\citet[W09;][]{West09}, and \citet{So98}. The C labels with the W09 data refer
to the different components of an IFU observation. 
}
\label{fig:compare} 
\end{figure*}

\citet{Te91} discovered M82's stellar bar with observations at infrared
wavelengths and estimated it to be approximately 1~kpc in length. 
Numerical simulations have shown that galaxy interactions are effective at
inducing galactic bars \citep[e.g.][]{No88, Mi98}, and the strong evidence for
past tidal interaction observed in M82 makes it an excellent testing ground for
theories of bar formation. Although it is difficult to observationally
determine the origin of M82's bar, the effects of its presence can be seen in
our stellar rotation curve as the symmetric, sharp features on either side of
the nucleus.  

The second bump at $\sim$500~pc may be the result of a super star cluster (seen
in the two-dimensional data) that fell on the slit.  This cluster, which was
identified as cluster `z' by \citet{Mc07}, likely dominates the flux
from this aperture, and its velocity excess with respect to the main rotation
curve likely indicates strong non-circular motion.  However, it is puzzling
that multiple points define the feature and that no other studies have seen
this second peak in the rotation curve.  Other possible explanations for the
feature are the presence of tightly wound spiral arms and patchy $K-$band
extinction. 

Our data contain evidence of a ``velocity reversal'' (velocity decrease) at
$\sim$1.0~kpc on both the eastern and western sides of the galaxy. \citet{Gr11}
highlights the existence of a 100~km~s$^{-1}$ velocity reversal of gas and/or
stars at $\sim$1.0~kpc in H$\delta$ and H$\epsilon$ stellar absorption on the
eastern side of the galaxy, as well as in ionized gas emission lines at visible
and blue wavelengths on the eastern and western sides; interestingly, the
feature is apparently absent in near-infrared Ca~II stellar absorption.
Although the velocity amplitude of our detection is approximately a factor of
10 less, the reversal is present in our data. Local extinction is not likely
to be the culprit, as the feature is symmetric with respect to the nucleus.
\citet{Gr11} investigates the hypothesis that the reversal is due to local
surface density enhancements associated with spiral arms \citep{Ma05}, but
finds inconsistencies with this picture. For instance, the absence of the
velocity reversal in near-infrared Ca~II stellar absorption suggests that it is
not inherent to the disk, and the blue-wavelength detections of the eastern
velocity reversal reach negative velocities, which cannot be explained from the
rotation of a stellar disk with embedded corotating spiral arms.  Nevertheless,
the detection of the reversal in \CO\ stellar absorption and the symmetry about
the nucleus indicate the feature is present in the stellar population and
implies it is due to a local density enhancement, such as spiral arms. 

The inset of Figure~\ref{fig:CO} compares our results for the inner region with
\citet{West09} and \citet{Mc93}. The agreement with \citet{West09} is excellent
outside the bar region. There is good agreement with \citet{Mc93}, as well as
some interesting differences. The sharp peaks just off of the nucleus are in
agreement on the eastern side but not on the western side. \citet{Mc93}
measured a larger amplitude for the western peak. Interestingly, the western
peak of our gas measurements (described next, and see Figure \ref{fig:compare})
is in better agreement with \citet{Mc93} than our stellar measurements.

\vspace{-.2cm}
\subsection{Gas Emission}

There are several prominent gas emission lines in our data that are
characteristic of starburst activity, most notably Br$\gamma$, He\,I, and H$_2$
(Figure~\ref{fig:spec}). The Br$\gamma$ and He\,I are recombination lines that
trace photoionized nebulae, and the H$_2$ is a rovibrational line that traces
shocked gas. In contrast to the observed strength of \bandhead\ relative to the
continuum, which decreases with distance from the nucleus, the emission lines
show an increase of intensity with distance to 350~pc radius and a rapid
decrease beyond 1~kpc radius. This is in qualitative agreement
with the intensity variations measured by \citet{Fo01}.

For each aperture listed in Table \ref{tbl:curve} with sufficient
signal-to-noise, we used the {\tt emsao} package within IRAF to measure the
velocity offset of each emission line relative to its rest wavelength
\citep{Cox}. We normalized the emission line rotation curves to the \bandhead\
velocity in an aperture $\sim$1.25$''$ west of the nucleus.
Figure~\ref{fig:compare} compares our gas and stellar rotation curves with the
nearly Keplerian gas dynamics measured in CO~(2,1) by \citet{So98}. The
emission lines used to measure the gas kinematics are only detected out to
1~kpc. 

On average, our measurements show that the gas moves with higher velocities
than the stars inside 1~kpc. This can be explained by forces from the stellar
bar, which tend to drive the gas towards the center of the galaxy.
Figure~\ref{fig:compare}a compares our gas measurements with our stellar
rotation curve and radio observations of CO~(2,1) by \citet{So98}. The dramatic
decline seen in \citet{So98} is not seen in the \bandhead\ rotation curve.
Figure~\ref{fig:compare}b compares gas measurements by \citet{Ca91},
\citet{Mc93}, \citet[W09;][]{West09}, and \citet{So98} with our measurements.
These data cover a wide wavelength range and are in fairly good agreement.
There is a large discrepancy in amplitude between \citet{So98} and the other
data. As the depth of an observation is a function of wavelength, these
discrepancies may be the result of optical depth variations. 

\section{Mass Distribution}

The mass of M82 has a much larger contribution from atomic and molecular gas
than seen in typical spiral galaxies. \citet{Yo84} estimate that between 30\%
and 40\% of the galaxy's mass is in the ISM and that the H$_2$/H~$\!\!$I mass
ratio is approximately 10:1. These authors postulate that the high gas fraction
may be due to infalling material that contributes to the available reservoir.
The estimated dynamical mass of M82 ranges from several times 10$^9 \rightarrow$
10$^{10}~M_{\odot}$ in the literature \citep[e.g.][]{Bu64, Cr78, Yo84, Go90,
Sofue92}; however, the substantial dust attenuation and non-circular motions
associated with tidal forces and the stellar bar make the estimation of the
mass challenging and uncertain.

\begin{figure}[h] 
\includegraphics[width=9.0cm]{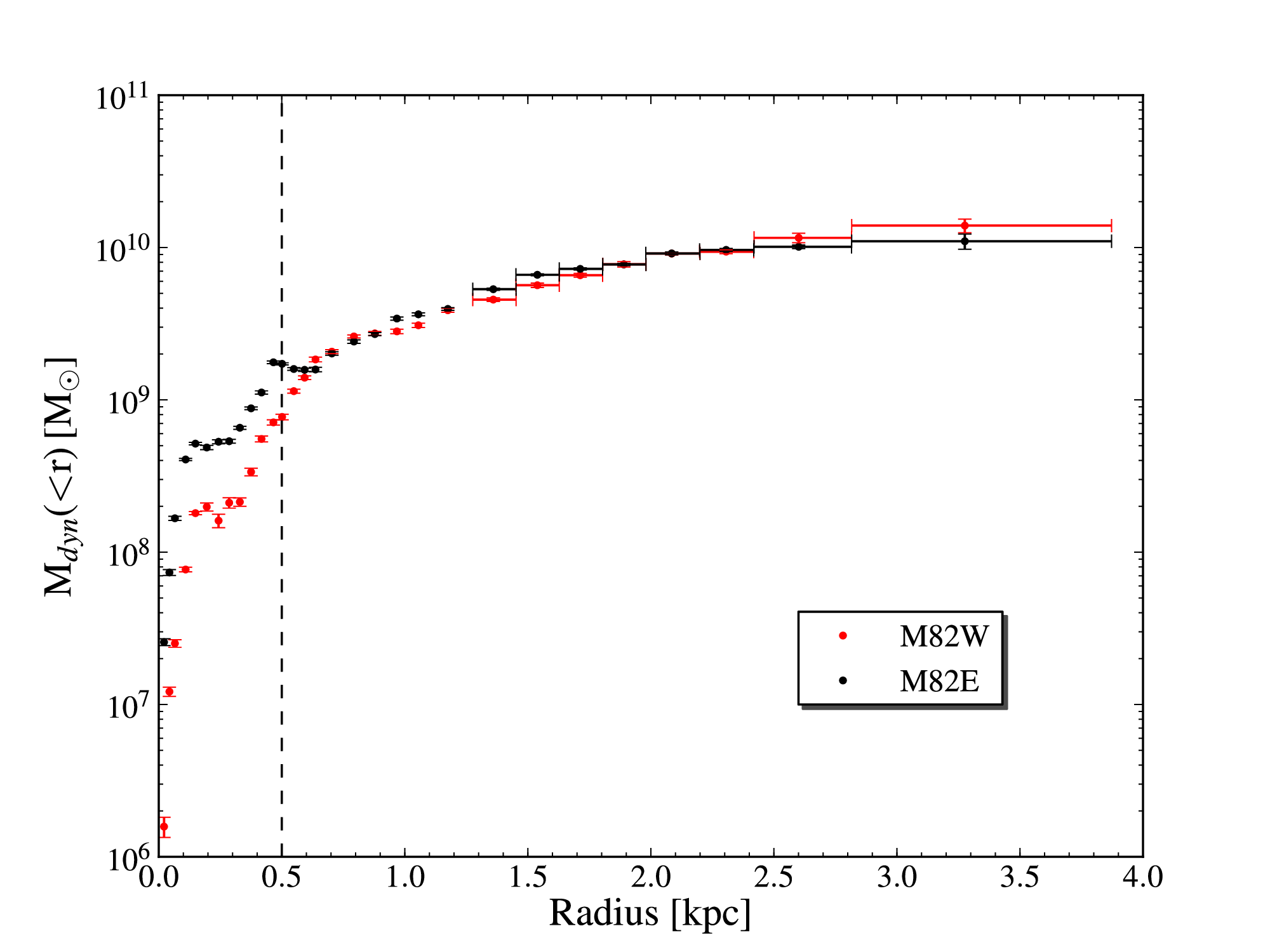}
\caption{
Mass profile derived from our stellar rotation curve. Circular motion was
assumed in the derivation. The horizontal error bars denote the aperture size
of the measurement. The apparent decrease in mass at small radii may be due to
the non-circular motion associated with the stellar bar. The superwind is
likely launched from the region inside the dashed line, where we estimate the
mass is $\lesssim 2\times 10^9~M_{\odot}$.
}
\label{fig:mass} 
\end{figure}

Figure~\ref{fig:mass} shows the mass profile based on our stellar kinematic
measurements. Similar to the studies mentioned above, we assumed circular
motion to calculate the enclosed mass at each point. While non-circular motions
are expected in the central region, and in particular in the bar region,
circular motion is a good assumption on large scales. At 4~kpc the mass
distribution appears to be asymptotically approaching
$\sim$10$^{10}~M_{\odot}$. This is likely the total mass of the galaxy, as
these measurements are based on stellar motions and extend out to nearly 4~kpc
from the nucleus. 

Circular motion is not a good assumption in the central region, where the bar
and perhaps other asymmetries in the potential produce non-circular motions.
Nevertheless, here we estimate the mass within 500~pc is  $\lesssim 2\times
10^9~M_{\odot}$, which is consistent with mass measurements of this region by
\citet{Yo84}, \citet{Go90}, and \cite{Gr02}. This mass estimate may be high by
a factor of two based on the relative mass distributions for M82E and M82W
shown in Figure~\ref{fig:mass}. 

\vspace{-.4cm}
\vspace{-.5cm}
\begin{figure}[h]
\includegraphics[width=9.0cm]{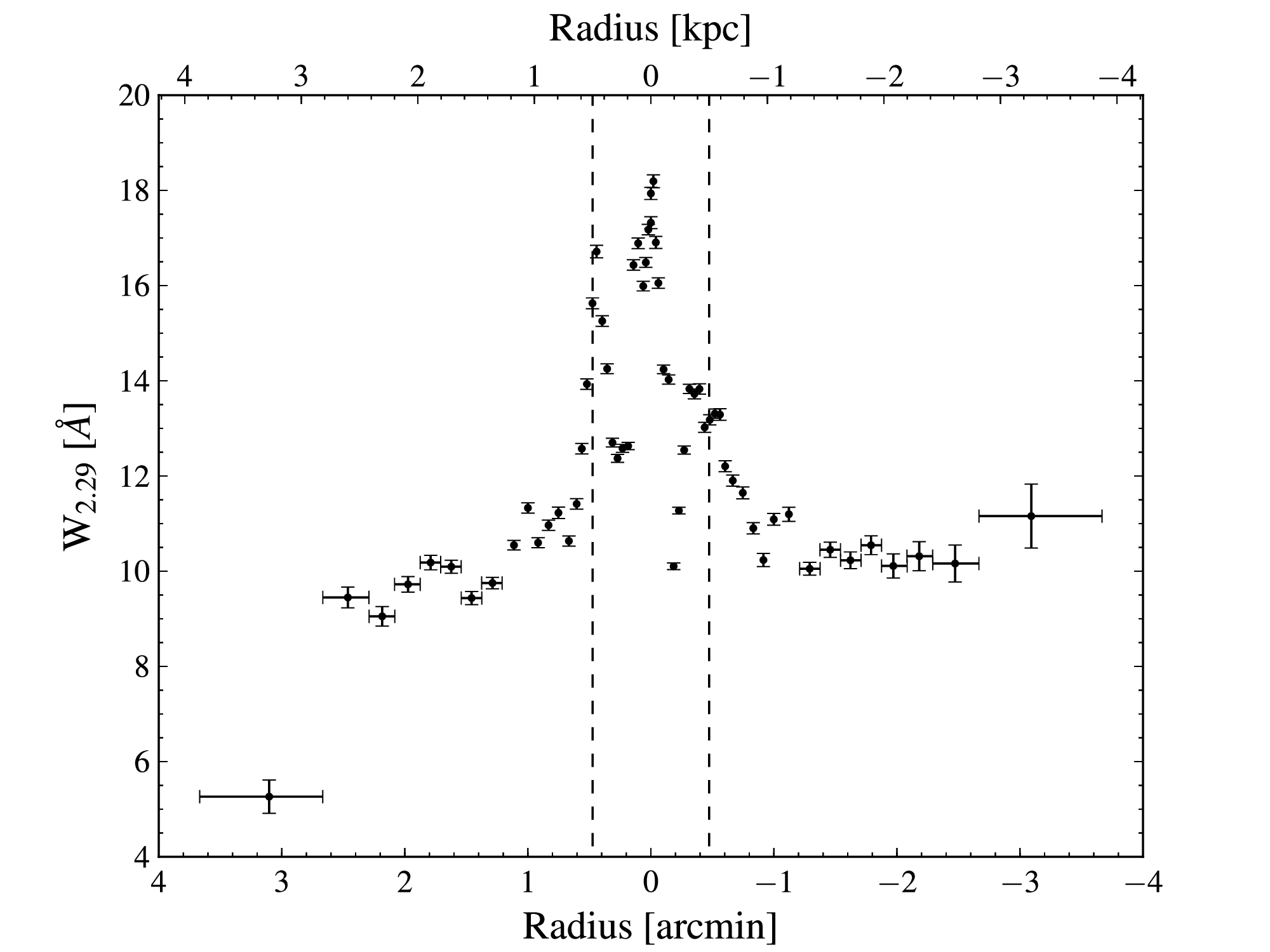}
\caption{
\ew\ as a function of distance from the center of the galaxy. The horizontal
error bars indicate the aperture size of the measurement. The peak in \ew\
inside the dashed region is indicative of a stellar population dominated by RSGs.
}
\label{fig:eqw}
\end{figure}
\vspace{-.5cm}

\section{Stellar Populations}

The CO absorption index in stars increases with increasing stellar luminosity,
decreasing effective temperature, and increasing metallicity \citep{Fr78,
Fr83}. Large CO indices are indicative of either a stellar population dominated
by RSGs or one dominated by old, metal-rich giants. These two possibilities
have led to much debate on the nature of M82's evolved stellar population,
particularly in the nucleus \citep{Fo01}.  Starburst models with RSG-dominated
nuclei, however, provide the best fit to the observed properties of M82
\citep{Ri93, Fo03}. 

Our data reveal very deep CO absorption in M82's starburst core.  In
Figure~\ref{fig:eqw}, we show the equivalent width of \bandhead\ (\ew) as a
function of radius.  The data clearly show a significant increase in \ew\
within 500~pc radius. Based on the diagnostic diagrams proposed by \citet{Or93}
and \citet{Ol95}, these results suggest that RSGs dominate the near-infrared
continuum throughout the starburst core. At yet smaller radii, \citet{Fo01}
found that \ew\ within the central 35~pc is characteristic of RSGs with
$T_{eff}=3600-4200$~K.

The dramatic contrast in \ew\ between the starburst core and the stellar disk
can be explained by the successive starburst model proposed by \citet{Fo03},
in which the disk and the starburst core are composed of two distinct
stellar populations.

\vspace{-.1cm}
\section{Summary} 

We have used longslit, $K-$band spectroscopy from the LBT to measure the
stellar and gas kinematics and study the mass and stellar population
distribution in M82. The primary points of this paper can be summarized as follows.

We used \CO\ stellar absorption to measure the rotation curve out to nearly
4~kpc on the eastern and western sides of the galaxy. While some previous
studies have interpreted H~$\!$I and CO emission-line position-velocity
diagrams as evidence for a declining rotation curve, our data show that it is
flat from 1~$-$~4~kpc (Figure~\ref{fig:CO}).  The rotation curve also contains
the signatures of the stellar bar and an apparent ``velocity reversal'' at
$\sim$1.0~kpc on either side of the galaxy that may be the manifestation of
symmetric spiral arms. 

We measured the rotation curve with the Br$\gamma$, He$\!$~I, and H$_2$
emission lines from the interstellar medium out to 1~kpc from the nucleus.
These gas rotation curves have velocities 10~$-$~50~km~s$^{-1}$ greater than
the stellar velocities (Figure~\ref{fig:compare}). The gas dynamics are
expected to be disrupted more than the stars by the presence of tides and
galactic winds, which is the likely origin of the stellar and gas velocity
differences.

We used the stellar absorption data to measure the mass distribution
(Figure~\ref{fig:mass}) and investigate the stellar populations as a function
of radius (Figure~\ref{fig:eqw}). We estimate the total dynamical mass of M82 is
$\sim10^{10}$~$M_{\odot}$.  The strong variation in \ew\ with radius clearly
indicates that the starburst is located within the central 500~pc radius.  The
superwind is likely launched from this region, and we estimate its mass is
$\lesssim 2\times 10^9$~$M_{\odot}$.  This mass estimate is somewhat uncertain
due to the presence of non-circular motions, but nevertheless provides an upper
limit to the gravitational potential of the starburst region.  

\vspace{-.3cm}
\acknowledgments 

We thank Mark Westmoquette for comments and supplying data for
Figures \ref{fig:CO} and \ref{fig:compare}. We also appreciate the helpful
comments from the referee. We are grateful to Jeff Blackburne and Jill Gerke for
taking the data as part of the OSU/RC queue. JPG is grateful for support from
an undergraduate research scholarship from the College of Arts and Sciences at
the Ohio State University. PM is grateful for support from the NSF via award
AST-0705170. TAT is supported in part by NASA grant NNXIOADOIG and an Alfred
P. Sloan fellowship.


\end{document}